# Eliminating nearfield coupling in dense high quality factor phase gradient metasurfaces

*Samuel Ameyaw[1] [†][*], Lin Lin[1,2][†], Bo Zhao[1], Hamish Carr Delgado[3] and Mark Lawrence[1][*]*

[1]Department of Electrical and Systems Engineering, Washington University in St. Louis, St. Louis, Missouri 63130, USA

[2]Department of Chemistry, Washington University in St. Louis, St. Louis, Missouri 63130, USA

[3]Department of Materials Science and Engineering, Stanford University, Stanford, California 94305, USA

*Corresponding author(s). E-mail(s): ameyaw@wustl.edu ; markl@wustl.edu

†These authors contributed equally to this work

## Abstract

High Q phase gradient metasurfaces are becoming promising elements for revolutionizing light manipulation but near-field coupling typically forces a trade-off between quality factor and resolution. Here, we present a strategy for not just reducing but eliminating coupling-based nonlocal effects in wave shaping metasurfaces composed of meta-pixels with arbitrarily long resonant lifetimes arranged with sub-diffraction spatial resolution. By working at a zero-coupling regime introduced by the interference between enhanced longitudinal and transverse electric fields, the tradeoff between Q and resolution no longer exists. Exploiting for wave shaping the ability to fully suppress coupling between high Q meta-atoms, we numerically demonstrate structurally uniform devices that produce beam-splitting to angles of ±53º and beam-steering to an angle of 33º with diffraction efficiencies over 90% via refractive index modulations of just $2\times10^{-6}$ and $7\times10^{-6}$, respectively. These are made possible by the underlying meta-structure supporting local dipole resonances with Q factors of 2.8 million and 0.87 million, respectively, arranged with a dense pixel pitch of $\lambda/1.6$. Extending the approach to structures with $\lambda/2.2$ resolution we also unlock full-field beam steering via index biasing of just $1\times10^{-4}$. To experimentally validate our scheme, we measure the incident angle dispersion of various high Q metasurface samples. The signature of a zero-coupling regime is discovered in the form of a sign flip in the angular dispersion with resonant wavelength. Aside from triangulating a perfect decoupling configuration, one of our fabricated nanofin-isolated metasurfaces with Q-factor >870 has a resonant wavelength that stays within the half linewidth for incident angles of -20º to 20º. The platform being introduced provides a route for densely arrayed high Q metasurfaces with independently addressable constituent meta-atoms, paving the way for highly efficient nonlinear and dynamic wavefront shaping.

## Introduction

Dielectric nanostructures have drawn a lot of interest within the photonics community because of their unique combination of dramatic light scattering and remarkably low loss. By engineering the size, shape, and/or composition of these structures, it's been shown that the enhanced scattering can be harnessed for exciting applications in free space communication, sensing, and computation[1–4]. Key to the precise control over optical scattering is the ability to interact strongly with specific wavelengths of light via Mie-like resonances[5]. Resonance allows the nanoscale objects to imitate long free space optical path lengths, giving access to a much wider range of scattered phase, amplitude, or polarization modulations[6–9]. Not only are these nanoparticles incredibly thin but as each has a subwavelength footprint, many of them can be arranged into a dense nonuniform array[10]. Known as a phase gradient metasurface, such an array can sculpt optical wavefronts with unprecedented precision by choosing the reflected or transmitted phase at each nanoscale pixel, often by locally tuning the difference between the resonant wavelength and illumination wavelength[11]. This tunability enables the shaping of wavefronts in ways that are not possible in conventional bulk optics enabling applications of dynamic beam steering devices[12], holograms[13], and compact imaging systems[14].

For next generation applications of phase gradient meta-optics, including AR/VR[15], LiDAR[6], quantum communication[16], and bio-sensing[17], the modest amplification of light-matter coupling provided by Mie resonance represents a significant impediment[18]. While omnidirectionally radiating to the far field makes Mie scatterers ideal point sources, access to such a large number of radiation channels causes their resonant modes to leak very quickly, characterized by a low quality factor (Q-factor), measured as the ratio between the resonant frequency and the resonant bandwidth[7]. As a low Q-factor corresponds to a broadband response and weak intensity enhancement, dielectric nanoparticles are not typically sensitive to small environmental changes and exhibit much weaker nonlinear signals compared to bulk crystals. This shortcoming makes them unsuitable for equipping phase gradient meta-optics with nonlinear or dynamic properties, at least within near term commercial technologies[19,20]. To address this challenge, nanostructures with very weak radiation leakage have been introduced as potential building blocks of nonlinear and programmable phase gradient metasurfaces[21,22,23]. Combining enhanced light-matter interactions and precise control over the phase distribution, these so-called high quality factor (Q-factor) metasurfaces promise a wide range of nonlinear and dynamic wave-shaping functionalities[24]. By dynamically controlling the resonant properties of the constituent high Q nanoantennas, for example, the phase distribution across a metasurface can be altered in real time, allowing for on-demand manipulation of transmitted, reflected, or diffracted light[25,26]. The strength of these devices originates from the fact that the light intensity circulating inside the resonant antenna grows as the linewidth decreases in direct proportion to Q. As a result of the enhanced light intensity, small illumination power is required for nonlinear functionality and a narrow linewidth allows for large modulation with a small external bias[27–31].

Aside from Q, another crucial metric by which to evaluate the performance of phase gradient metasurfaces is resolution, as it directly influences the accuracy achievable in wavefront shaping as well as the accessible angular field of view, or numerical aperture. The resolution of phase gradient metasurfaces is predominantly limited by nearfield coupling between neighboring meta-atoms, which prevents independent light scattering and generates cross-talk[32]. Nearfield coupling can be reduced by shrinking the evanescent decay lengths of the individual modes, thereby reducing the mode overlap[33]. However, the diffraction limit ultimately prevents dielectric resonances from being localized below the wavelength scale[34]. Nevertheless, wavelength scale dielectric metasurfaces have been successfully applied to the development of highly efficient and high performance flat optical lenses[35], holograms [36]. While narrow linewidths and long decay times are indeed advantageous properties associated with high Q resonance, they also further compound the issue of nearfield coupling. Being extremely sensitive to their environment, when placed close together high Q resonators become susceptible to even tiny nearfield interactions, spoiling their ability to produce independent phase delays[37]. There seems therefore to be a trade-off in the design of local metasurfaces between resonant enhancement and wave-shaping resolution. The higher the Q-factor, and thus the stronger the light matter coupling, the further the meta-atoms must be spaced apart, leading to less precise wavefront engineering.

In this study, we demonstrate that polarization-based interference can eliminate the trade-off between meta-atom quality factor and meta-atom spacing, unlocking high resolution wave-shaping via tailored refractive index patterns with infinitesimal modulation depths. Specifically, we reveal that the longitudinal evanescent electric field between guided mode resonant silicon nanoantennas can be enhanced if the intervening space is filled with structurally birefringent silicon nanofins. After careful tuning of the resonant wavelength, nearest neighbor coupling via the longitudinal electric field can be brought into balance with, and therefore perfectly cancel, nearest neighbor coupling via the transverse electric field, as shown in Figure 1a. As proof-of-principle for the power of this coupling suppression, we simulate high angle beam-splitting ($\pm 53°$) and beam-steering ($33°$) with diffraction efficiencies exceeding 90% using index contrasts as low as $2\times 10^{-6}$. This is made possible by embedding independent resonators with Q as high as 2.8 million in metasurfaces with subwavelength pitch of $\lambda/1.6$. Finally, we experimentally verify our scheme by measuring a sign flip in the incident angle dispersion for a series of geometrically swept metasurfaces. Aside from observing a clear signature of decoupling, we realize a sample with Q>870 that maintains its resonant wavelength within the half linewidth over a 40° field of view, opening possibilities for narrowband filtering of high-resolution images.

## Results

As illustrated schematically in the inset of Figure 1b, our metasurface design consists of meta-atoms made from silicon nanobeams sitting on top of a $SiO_2$ substrate and periodically perturbed with small notches to break translational symmetry and form guided mode standing waves[38]. As these standing waves remain perfectly trapped inside nanobeams without notches, due to a momentum mismatch with free space, we can engineer dipolar scattering resonances with arbitrarily large Q-factors by using arbitrarily small notch depths or widths[39,40]. To turn these dipolar guided mode resonators (DGMRs) into reflective phase pixels suitable for wavefront shaping, we place a metal ground plane at the back of the $SiO_2$. After aligning the illumination wavelength to the DGMR, the reflected phase can be tuned from 0-2π by adding a slight modification Δn to the refractive index of silicon inside a nanobeam. This modification could be achieved with various mechanisms including thermo-optics[41], the Pockels effect[42], the nonlinear Kerr effect[43], or simply structural tuning[37], but we will not worry about those details here. By choosing different Δn's in three neighboring meta-atoms, a phase gradient should be imparted on an incident plane wave, steering the reflected wave away from the normal direction. However, in Figure 1b no such steering can be seen. This is because nearfield coupling breaks the simple relationship between the refractive index of a meta-atom and the phase reflected from it.

**Figure 1.** Using interference between two polarization pathways with opposite symmetries to eliminate near field coupling in densely packed high DGMRs allowing efficient high Q beamsteering.

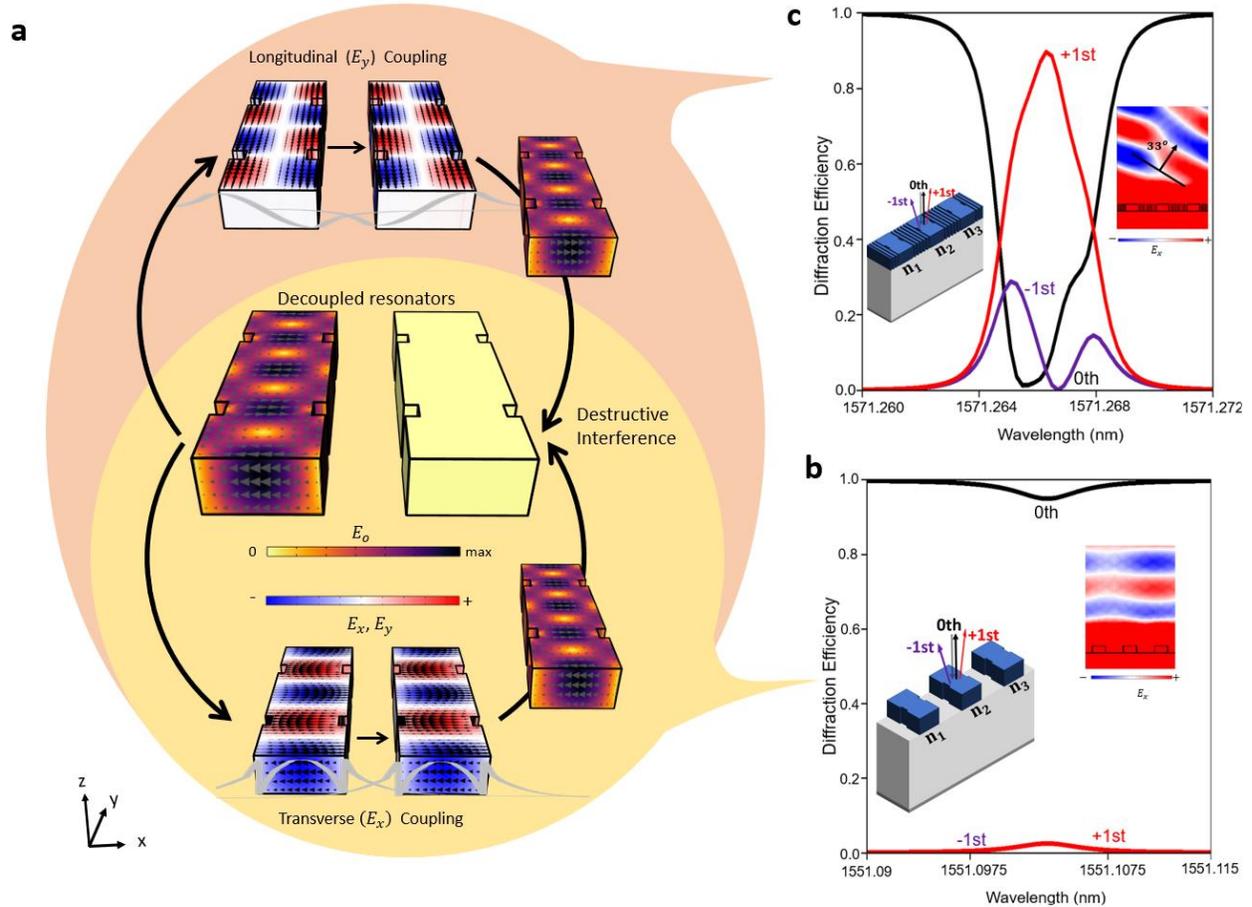

a) Schematic showing the decoupling mechanism in high Q DGMRs. Destructive interference between the two polarization pathways with opposite symmetries allows the decoupling of the DGMRs. (b) Metasurfaces without fins show negligible diffraction efficiency for a Q=922,590, $\Delta n=1\times10^{-3}$ with notch dimension 10nm×30nm. Left inset: schematic illustrating diffraction configuration for a beam steering device without fins (In all schematics, unless otherwise specified, blue represents silicon, light grey denotes $SiO_2$, dark grey indicates PEC boundary, $n_1 = n - \Delta n/2, n_2 = n, n_3 = n + \Delta n/2$ are the respective refractive index of the silicon blocks). Right inset: electric field profile of coupled metasurface at $\lambda = 1551.103 nm$  c) Metasurface with extreme decoupling shows high diffraction efficiency for small index bias of $\Delta n=7\times10^{-6}$ and Q = 870,634 with notch dimension 10nm × 30nm. Left inset: schematic illustrating diffraction configuration for a beam steering device with fins. Right inset: electric field profile of decoupled metasurface at $\lambda = 1571.266 nm$ with a beam steering angle of 33º.

Whereas in Figure 1c, nearfield coupling has been eliminated, restoring the phase gradient and producing efficient steering to an angle of 33° using elements with Q~870,634, 30x larger than has been shown before[44].

Building upon results in Figure 1, it is evident that high Q DGMRs, despite their cross-sections being as localized as Mie modes, are much more sensitive to near-field coupling. To understand this, we need to recognize that coupling strength physically represents the rate at which energy can pass from one structure to another. The absolute rate is, therefore, less important than the relative rate when compared to how fast energy can leak out of either structure, radiating to free space. For low-Q Mie scattering, light spends so little time in the nanostructures that the nearfield overlap must be fairly large in order to allow for a significant fraction of the incident energy to make it across. For DGMRs on the other hand, the higher the Q-factor, the longer light remains trapped inside the initial resonator, allowing for significant energy transfer between neighbors even for vanishingly small overlap of weak evanescent tails. To quantify the balance between far-field decay and nearfield coupling for two identical DGMR nanoantennas placed close together, we need to analyze the formation of hybrid modes consisting of symmetric and anti-symmetric oscillations of the single antenna modes with two new resonant frequencies $\omega_s$ and $\omega_{as}$. As long as the frequency splitting $|\omega_s-\omega_{as}|$ remains smaller than the individual antenna linewidths, shifts in the individual resonant wavelengths can produce arbitrary scattered phase differences between the pair[45]. However, if the linewidths drop below the splitting, the scattered phases become insensitive to the individual resonant wavelengths, being instead locked to the hybrid modes. This is the root of the trade-off between resonant enhancement, or Q factor, and phase front resolution in phase gradient metasurfaces.

**Figure 2**. Finding of hybrid mode crossing by tuning the resonant frequency of nanofin-isolated metasurfaces.

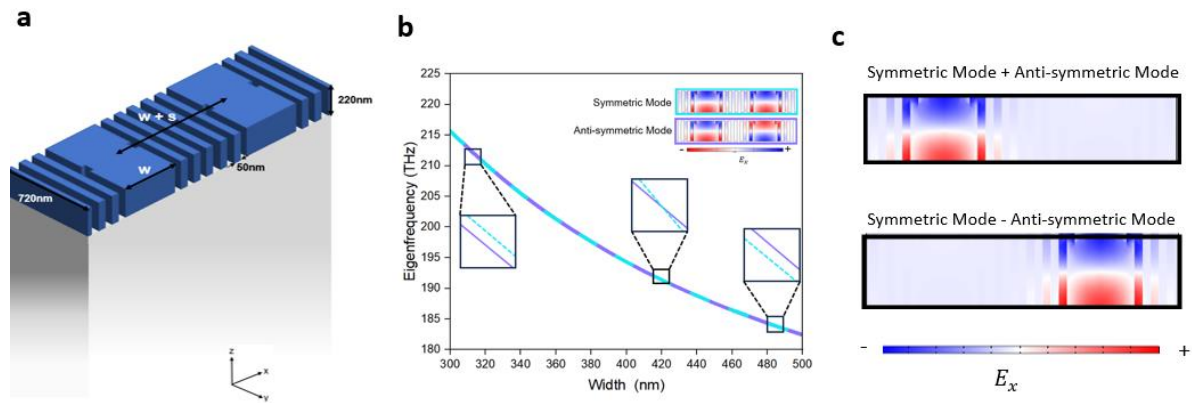

a) Schematic metasurface with anisotropic fins with notch dimension 50nm×100nm, w=420nm, s=550nm b) Zero-coupling regime, where the symmetric and anti-symmetric modes cross when the width of Si nanoantenna is varied inset top right: Symmetric and antisymmetric mode. c) At the crossing point (Si nanoantenna width of 420nm), because the symmetric and antisymmetric modes have the same frequency, they can be superimposed in various ways to achieve a well-defined localized mode in each resonator.

Our design strategy for eliminating the trade-off is illustrated in Figure 2a. Placing thin silicon nanofins between the DGMR nanoantennas modifies the evanescent fields by mimicking a highly birefringent material. Such birefringence has been shown to decrease the decay length of transverse electric fields[33,46], but while the corresponding decrease in coupling strength for a given antenna spacing raises the maximum Q factor, the trade-off persists. Inspired by recent work on the elimination of waveguide coupling[47], we reveal a much more dramatic consequence of the fins in Figure 2b. When varying the widths of the nanoantennas, not only do the hybrid mode frequencies change, but for a particular width the symmetric Eigen-frequency crosses the anti-symmetric Eigen-frequency. At such a crossing, the coupling strength is not merely reduced but becomes identically zero. This can be seen by exploiting the degeneracy of the hybrid modes and linearly combining the symmetric and anti-symmetric fields, shown in the inset of Figure 2b, revealing a new set of Eigenmodes located in one nanoantenna or the other, shown in Figure 2c.

Instead of decreasing the total overlap between nearest neighbors, the decoupling singularity emerges after increasing only the longitudinal field overlap. As shown in Figure 1a, a DGMR consists of two orthogonal polarization components: the longitudinal and transverse electric fields. These represent two separate pathways for coupling to adjacent resonators. The transversely polarized electric field pattern is symmetric with respect to the y-z plane at the center of the nanostructure and so generates in-phase oscillations within neighboring structures, corresponding to a positive coupling coefficient. Conversely, with a distribution that transforms anti-symmetrically about the y-z plane at the center of the nanostructure, the longitudinally polarized electric field excites neighboring resonances that oscillate with a π phase delay, corresponding to a negative coupling coefficient. This π phase difference therefore causes the two coupling routes to destructively interfere. Usually, such interference is insignificant because one polarization, and therefore one coupling pathway, is much stronger than the other. This is the case in Figure 2b, where the transverse field dominates. But, after introducing anisotropic fins, the longitudinal evanescent electric field is enhanced while the transverse evanescent electric field is simultaneously suppressed. Varying the nanobeam widths then serves to fine tune the balance between these two polarizations until at a width of 420 nm they perfectly cancel out leading to a zero-coupling regime and a diverging coupling time[47].

To show the benefit of this zero-coupling regime in the decoupled nanoantennas for diffraction, we place a periodic array of these nanoantennas supporting DGMR placed above a perfect electric conductor (PEC), the PEC acts as a perfect mirror such that light impinging on the metasurface is reflected. The reflected wave can be engineered by spatially modulating the reflected phase. This spatial modulation of the reflected phase is achieved by independently tuning the resonant frequencies of the DGMR nanoantennas. We can achieve a variety of transfer functions, such as optical beam steerers, by varying the reflected phase between 0 and 2π.

Since it is easier to test decoupling and efficient diffraction with only one varying parameter (i.e. Δn) in beam splitting, we first demonstrate binary phase grating beam-splitters shown in Figure 3a-b, but this idea extends to any transfer functions, including beam-steerer shown in Figure 1c. Given that structural changes (notch dimensions) are made to the nanoantennas to manipulate the Q which affects $\omega_s$ and $\omega_{as}$, for ease of simulation, the average refractive indices of the nanoantennas are slightly tuned to ensure each device is working close to the crossing (Supplementary Figure 3).

**Figure 3.** Coupled dipole model predicts arbitrary high Q metasurface beamsplitters with near perfect diffraction when close to the crossing.

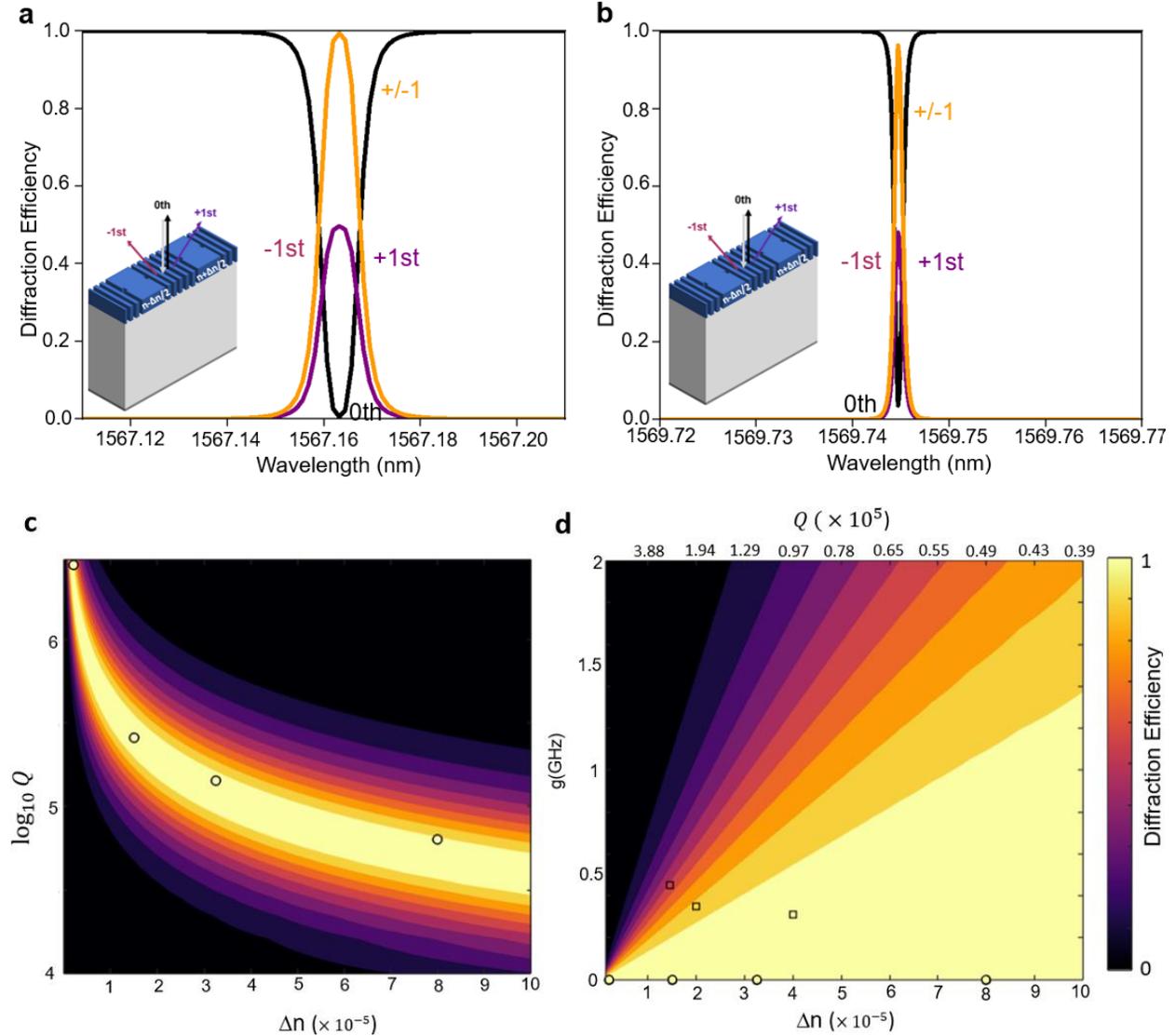

a) Diffraction efficiency plot for a Q=257,062, Δn=1.5×$10^{-5}$ with notch dimensions 10nmx40nm. inset: schematic of a reflective high Q metasurface with anisotropic fins in diffraction configuration b) Diffraction efficiency plot for Q=2,800,610 Δn=2 × $10^{-6}$ and with notch dimensions 10nmx10nm. inset: schematic of a reflective high Q metasurface with anisotropic fins in diffraction configuration. (c,d) Color map of Analytical Coupled dipole model with c)g=0, the black circle rings represent numerical simulations on the crossing(specifically the top two rings represent data of Figure 3b and Figure 3a respectively). b) g is set to a finite value ranging from 0 and to 2GHz, the black circle rings represent numerical simulations at the crossing whilst the black square rings represent numerical simulations away from the crossing.

By applying a subtle refractive index bias to a pair of the silicon beams in the binary phase grating, with opposite signs for neighboring beams, we achieved a diffraction efficiency of 99% with a Q of 257,062 and $\Delta n=1.5\times10^{-5}$ as shown in Figure 3a. Furthermore, the same structure but different notch dimensions is realized to function as a perfect beam-splitter with a diffraction efficiency of 96%, Q = 2.8 million, and a small index bias of $2\times10^{-6}$ as shown in Figure 3b. The presented data emphasizes the benefits of high Q metasurfaces as only a small index variation as low as $2\times10^{-6}$ is needed for perfect diffraction. The small index biases can be translated as negligible temperature variation in thermo-optic applications and small DC electric fields in electro-optic applications, for example, indicating that magnified modulation efficiency can be realized. More importantly, we believe the Q of these shown devices can be much higher, but such structures become increasingly difficult to model.

To explain how our platform can not only decrease but entirely eliminate nearfield coupling, enabling arbitrarily high Q and efficient diffraction simultaneously, we analytically describe the optical response of metasurface beam-splitters via a coupled dipole model. The Lorentz resonance model for two coupled dipole oscillators can be expressed as[48]

$$\begin{bmatrix} \delta_1 + i\gamma_1 & g \\ g & \delta_2 + i\gamma_2 \end{bmatrix} \begin{bmatrix} p_1 \\ p_2 \end{bmatrix} = \begin{bmatrix} g_1 E_{inc1} \\ g_2 E_{inc2} \end{bmatrix} \quad (1)$$

where g is the coupling term, which is chosen as a constant value in modeling, $\delta_1, \delta_2$ is the real part of the resonance of each of the dipoles and $\gamma_1, \gamma_2$ represents the decay rates of each of the dipoles respectively. Working with a normalized and uniform incident field, and assuming the oscillators have identical radiative loss rates that remain insensitive to small index changes, $g_1 E_{inc1} = g_2 E_{inc2} = 1, \gamma_1 = \gamma_2 = \gamma$, $\delta_1 = \omega - \alpha_1, \delta_2 = \omega - \alpha_2$,

$\omega = (\alpha_1 + \alpha_2)/2$. The two dipole moments $p_1$ and $p_2$ can then be written as

$$p_1 = \frac{-g+(\omega-\alpha_2+i\gamma)}{(\omega-\alpha_2+i\gamma)(\omega-\alpha_1+i\gamma)-g^2} \quad (2)$$

$$p_2 = \frac{-g+(\omega-\alpha_1+i\gamma)}{(\omega-\alpha_2+i\gamma)(\omega-\alpha_1+i\gamma)-g^2} \quad (3)$$

Our chosen relationship between $\alpha_1$, $\alpha_2$, and $\omega$ represents a pair of resonant frequencies shifted anti-symmetrically about the driving frequency. In the numerical simulations in Figures 3a-b, $\alpha_1$ and $\alpha_2$ depend linearly on the refractive index of silicon used in the respective nanostructures. We therefore use curve fitting to eigenmode calculations for symmetrically biased metasurfaces to find $\alpha_{1/2}$ and g as a function of the common index adjustment $\Delta n_1 = \Delta n_2$ (Supplementary Note 3). It is worth noting that g only drops to zero for a specific common index, and so one could worry about coupling arising after introducing an index contrast ($\Delta n = n_1 - n_2$) between neighboring dipoles. Interestingly, we see very good agreement between the simulation and our analytical model if we choose g based on the eigenmode splitting with common index equal to the average of the dissimilar indices, as detailed in the Supplementary Information Figure S4. A complex valued Fourier series was then used to calculate the strengths of the different diffraction orders contained in the secondary scattering from the periodic binary-phase meta-gratings[49]. Having tuned the average index to the crossing, Figure 3c shows that the Δn needed to achieve perfect beam splitting drops linearly with increasing dipole Q-factor. This trend is verified by the alignment between the simulation data of the meta-structure (represented by black circles) and the analytical model, both reaching a high diffraction efficiency of 99%. This outcome underscores the feasibility of achieving arbitrarily high Q and efficient diffraction concurrently at the zero-coupling regime. In Figure 3d, we explore what happens away from the crossing by varying the average index and Q-factor while tracking the peak diffraction efficiency, with the corresponding index contrast needed to reach the peak labeled at the bottom. In both the analytical model and the simulation (represented by black squares), the maximum diffraction efficiency attainable clearly deteriorates with increasing g for a fixed Q and with increasing Q for fixed g. Nevertheless, for any Q there remains a range of g values that can be tolerated while maintaining perfect diffraction control. So while mode overlap may be inevitable, this finding emphasizes the fact that, instead of meta-atom cross-talk, fabrication imperfections and absorption that limit experimental Q-factors are the only real barrier to minimizing the bias strengths required for high resolution wavefront shaping.

Advancing our decoupling principle from beam-splitting to beam-steering, the universality of our strategy is also demonstrated. A phase gradient metasurface consisting of decoupled resonant meta-atoms is realized with a pixel pitch of $\lambda/1.6$ enabling large diffraction angles and strong diffraction with a Q factor of 870,634. The beam-steering device is meticulously assembled using an array of resonant nanoantennas, engineered to establish a linear phase gradient ranging from 0 to $2\pi$ over the metasurface supercell. This is accomplished by configuring each unit cell to exhibit a phase shift of $2\pi/m$, where 'm' denotes the number of nanoantenna units within each supercell, thereby facilitating a complete $2\pi$ phase variation. This design allows for the steering of normal incident light to a predetermined angle in accordance with the generalized Snell's law as[50] $\theta_r = \arcsin(\lambda/n_i p)$ where $\lambda$ is the resonant wavelength, $n_i$ is the incident refractive index, $p$ being the supercell size. As in Figure 1c, with three nanoantennas per phase gradient unit cell and parameters p=2910nm, and $\lambda$=1571.266nm, the metasurfaces's high Q nature necessitates only a diminishing refractive index bias of $7\times10^{-6}$ is needed for efficient diffraction. This results in the 90% normal incident light being efficiently directed to the +1st diffraction order at a 33° angle. Conversely, a phase gradient metasurface with a comparable Q factor of 992,590 yet lacking the anisotropic fins, demonstrates negligible diffraction, despite an even higher index variation of $1\times10^{-3}$. This lack of efficiency is attributed to the substantial coupling between adjacent nanoantennas, emphasizing the crucial role of reducing coupling to achieve effective steering.

Having observed that the only requirement for perfect decoupling is the existence of a hybrid mode crossing and having verified that at a crossing arbitrarily high Q metasurfaces with $\lambda/1.6$ pixel pitch can near-perfectly diffract to angles up to 33°, now we seek to extend the angular range available to our high Q wave shaping platform. Using the same formalism, we look for a zero-coupling regime after bringing the nanoresonators closer together to produce a deeper subwavelength phase front shaping resolution of $\lambda/2.2$. To ensure that a high Q phase gradient metasurface with a resolution $\lambda/2.2$ can be realized, this critical condition should be satisfied: A crossing should be found. Sweeping the height of the silicon structures, rather than the width, we find a crossing between symmetric and anti-symmetric modes at an increased height of 383nm, as seen in Figure 4a-b. With a Q of 68,797 and an index bias Δn=$1\times10^{-4}$, Figure 4c shows beam-steering for a subwavelength resolution $\lambda/2.2$ metasurface, with normal incident light being deflected to the +1st diffraction order with an efficiency of 85% at a steering angle of 47°. This deep subwavelength $\lambda/2.2$ phase gradient device enables steering of light across a broad field of view, exceeding 90°, thus providing access to a wide range of angles. In fact, a 47° half angle is only a limit for periodic metasurface settings. When driven aperiodically, any antenna array with pitch $< \lambda/2$ can produce sidelobe-free steering across the full 180-degree field of view. This means that having extended perfect meta-atom decoupling to $\lambda/2.2$, we have in-principle unlocked full field beam-steering requiring vanishingly small reprogramming power.

**Figure 4.** Decoupled metal-backed high-Q metasurfaces beam steering device with a subwavelength resolution of $\lambda/2.2$

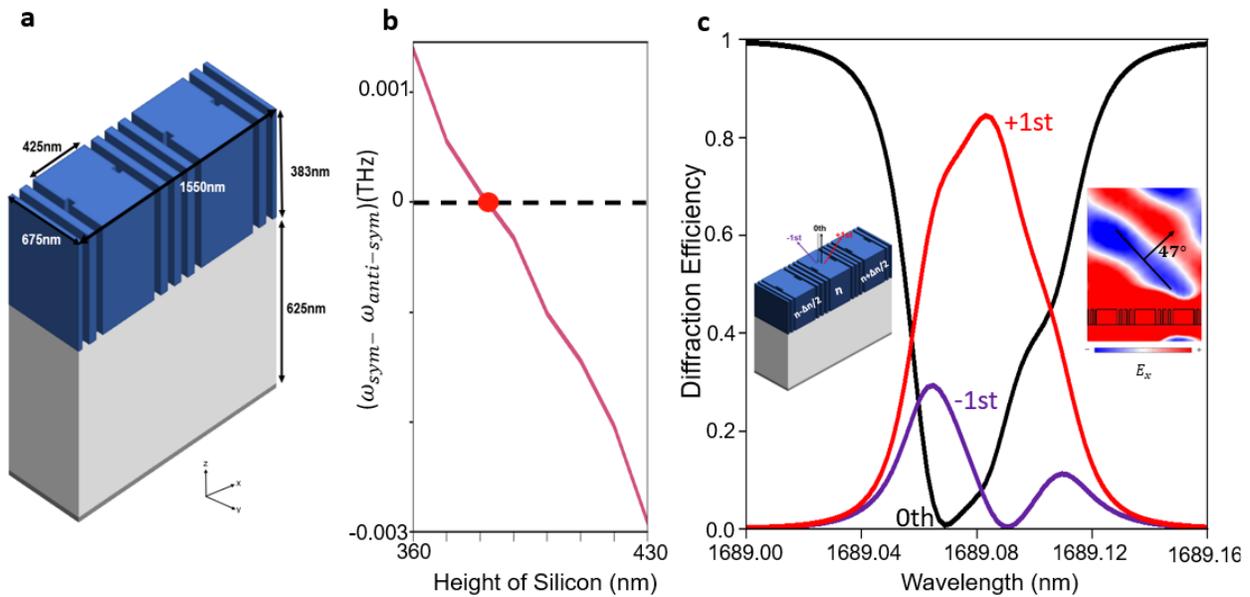

a) Schematic of a reflective high Q metasurface with a resolution of $\lambda/2.2$  b) Plot of difference between symmetric and antisymmetric eigenfrequency and variation of Silicon height c) Diffraction efficiency plot achieved with an index bias $\Delta n=1\times10^{-4}$ and Q = 68,797 and with notch dimensions 10nm×100nm. Inset left: schematic showing a reflective high Q metasurface with anisotropic fins in diffraction configuration. Inset right: Electric field profile of decoupled metasurface at $\lambda = 1689.08 nm$ showing beam steering at an angle of 47°.

Above we have focused on achieving high resolution wavefront shaping when illuminating non-uniform metasurfaces with a normally incident plane wave. Another important consequence of near-field coupling occurs even for uniform metasurfaces. When the incident wave is tilted away from normal, the corresponding phase gradient introduces an incident angle dependent excitation phase difference across neighboring resonators. Occurring with a fixed phase, nearfield coupling then introduces incident angle dependent interference which causes the resonant wavelength to shift when increasing the incident angle[51]. A similar criterion applies here to the case of pixelated diffraction optics above. For a given Q-factor, the maximum incident angle a device can accept is set by the coupling based spectral shift that matches the resonant linewidth. Higher Q metasurfaces therefore typically operate with a smaller numerical aperture. This is a second trade-off the scheme we are presenting can overcome. Leveraging the zero-coupling regime, we numerically show near-perfect angle independence. We operate our realized device in transmission without the PEC backing.

After sweeping the notch period "p" a crossing is found when p = 671nm, as shown in Figure 5a. Confirming that nearest neighbor coupling leads to incident angle dispersion, the cyan curve in Figure 5b, representing a high Q metasurface without nanofins, reveals a drastic blue shift of the resonant wavelength of over 100x its FWHM as the incident angle increases from 0º to 45º. When nanofins are added, the incident angle dispersion becomes very sensitive to changes in p. For metasurfaces including nano-fins with p=630nm and p=700nm, residual coupling causes a spectral shift 2-3× the FWHM, as expected from the finite mode splitting in Figure 5a.

**Figure 5.** Numerical simulation demonstrating High Q metasurface with near perfect angle independence and a signature of a zero-coupling regime.

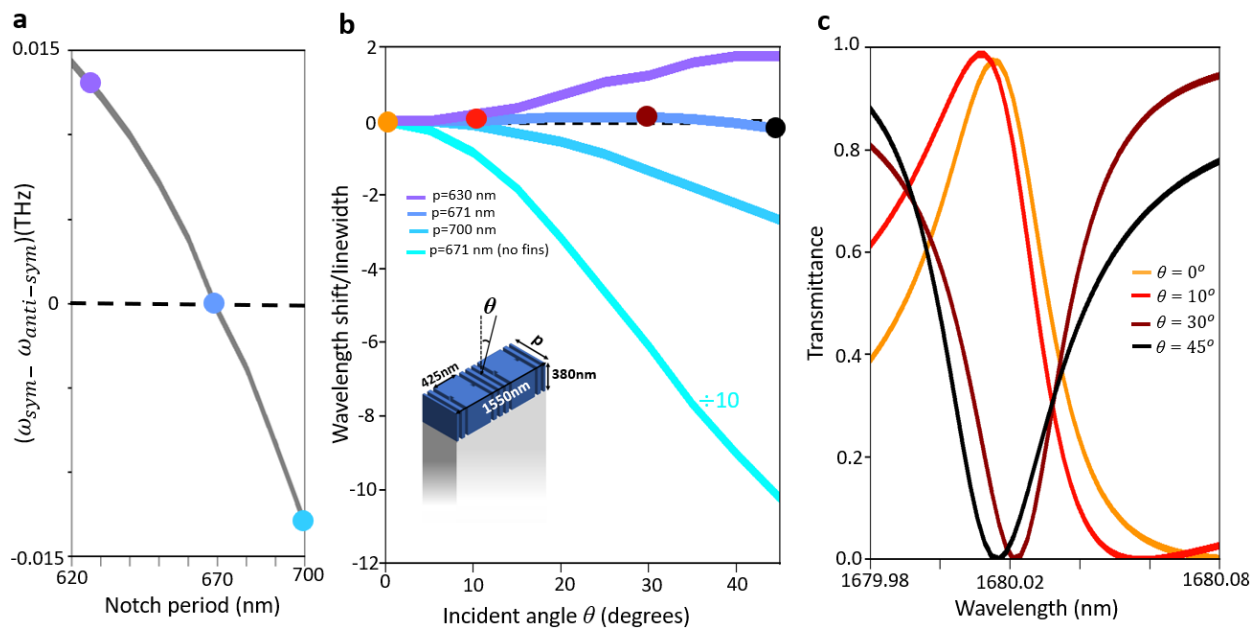

a) Plot of eigenfrequency as a variation of notch period. b) normalized wavelengths shift as a function of incident angle. The cyan curve represents a metasurface without fins with a Q of 67,520 and a period 671nm. The turquoise curve represents a metasurface with fins away from the crossing with a notch period=700nm and Q=45,429. The navy blue curve represents a metasurface on the crossing with a notch period =671nm and Q=44,876. The light violet curve represents the metasurface with fins away from the crossing with a notch period=630 nm and Q=46,549. Inset: Schematic of high Q angle independent metasurface. c) Transmission spectrum for metasurface with a fin at the crossing with a notch period =671nm.

Importantly however, these shifts have different signs, with the p=630nm sample's resonant wavelength increasing with incident angle while the p=700nm sample's resonant wavelength decreases. This will be crucial below for experimentally triangulating the crossing. Choosing p=671nm, we see from the navy blue curve in Figure 5b that incident angle dispersion has been almost entirely suppressed, the resonant wavelength shift remaining within the initial FWHM for the range 0º to 45º which indicates that the nanofins have fully decoupled the DGMRs. Furthermore, we explore the transmittance characteristics of the decoupled structure. From Figure 5c, the wavelength shifts stay within the FWHM which agrees with Figure 5b however, the transmission spectrum changes from a peak for θ=0º,10º to a dip for θ=30º,45º. This change is a result of the interference between the guided mode resonance and the background Mie resonances.

By tuning the resonant wavelength via the notch period we have demonstrated numerically in Figure 5b that the metasurface transmission spectra reveal a signature of the zero-coupling regime in the form of a sign flip of the incident angle dispersion (blue shift to red shift). Next, we experimentally confirm this numerical prediction by looking for a sign flip of the incident angle dispersion for a series of metasurfaces fabricated with different notch periods. Standard electron beam lithography and reactive ion etching has been used to pattern metasurfaces from a 350nm thick silicon film atop a sapphire substrate. Figures 6a-b show scanning electron microscope (SEM) images of our fabricated samples, with (Figure 6a) and without (Figure 6b) nanofins, with notch period p and nanoantenna width w annotated. We optically characterize the fabricated devices by illuminating them with a TE-polarized supercontinuum laser from the substrate side with the sample mount tilted to realize varying incident angles, ranging from 0º to 40º (Limited by the thickness of our sample holder). Transmission spectra are subsequently recorded using a home-built microscope coupled grating spectrometer (Supplementary Figure 5). Figure 6 c-d reveal how the transmittances of nano-fin metasurfaces with two different notch periods evolve with incident angle increasing from 0º to 40º. In Figure 6c, at normal incidence, the resonance wavelength for the p = 670nm metasurface is observed at 1547.86nm with a fitted Q of approximately 553. As the incidence angle is changed from 0° to 40°, the resonant wavelength blue shifts but stays within the FWHM up to 30°. Notably, between 20° to 40°, the resonant peak transitions to a dip, indicative of interference between the guided mode resonance and the background. Figure 6d shows at normal incidence, the resonance wavelength for the p = 530nm metasurface is observed at 1341.84nm with a fitted Q of approximately 872. As the incidence angle is changed from 0° to 40°, the resonant wavelength red shifts and overall stays within the FWHM for up to 80º (-40º to 40º).

Figure 6e compares the resonant wavelengths, extracted by fitting Lorentzian resonance curves to the spectra, of three samples. From the cyan curve we see that without nano fins there is a large spectral shift corresponding to almost 3 linewidths as the incidence angle approaches 40º. Although this is not as dramatic as shown in Figure 5b because we are dealing with lower Q-factors, the consequence of coupling still causes the metasurface with a moderate Q~1165 to be highly sensitive to the incident angle. Not only do the purple and blue curves in Figure 6c, for the p=530nm and p=670nm samples with nano-fins, exhibit strong suppression of coupling, with resonant wavelength shifts staying within the half linewidth up to 40º (i.e -20º to 20º). The two dispersion directions namely the red shifting and blue shifting of the transmittance spectra of the fin-isolated metasurfaces shown in Figure 6 c, d is an indication that there exists a perfect decoupling condition in between notch periods of 530nm and 670nm as similarly shown in the numerical simulation in Figure 5b.

**Figure 6**: Experimental verification of the signature of a zero-coupling regime in high Q nanofin metasurface.

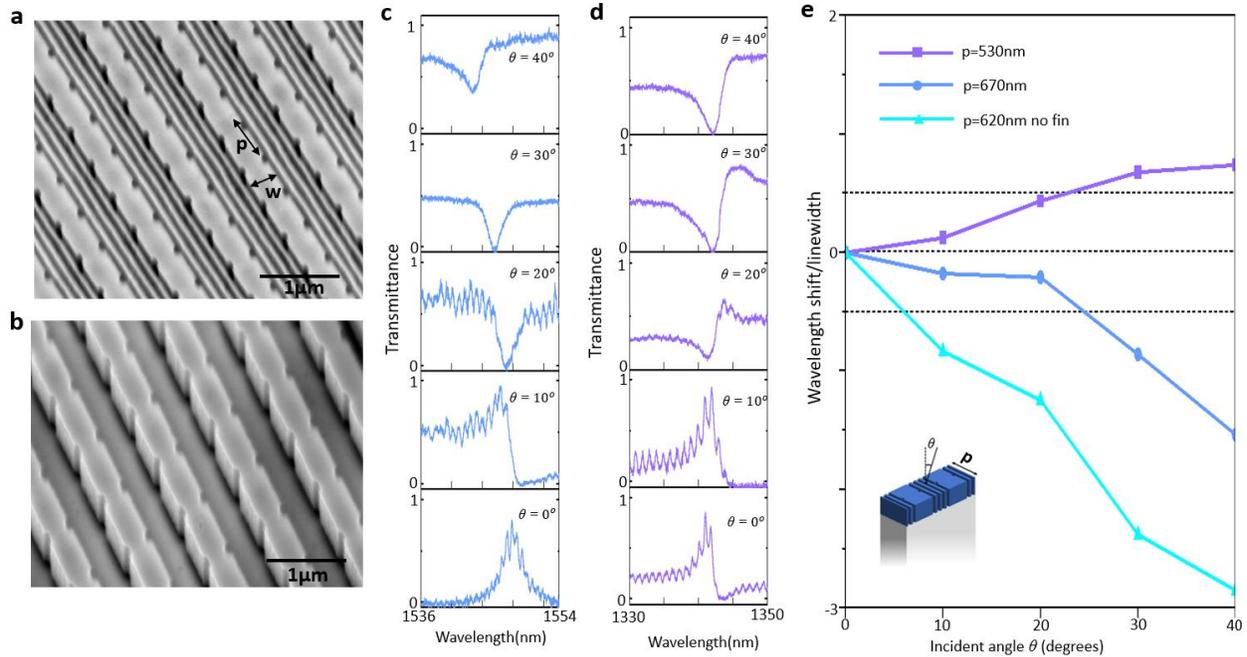

a)Angle SEM image of fabricated High Q metasurfaces a)with fins, with w=350nm,p=530nm, fins size=55nm, gab between fins= 70nm, gab between bar and fin =70nm, and notch dimensions 50nm × 75nm.b) without fin w=350nm,p=620nm c,d) Measured transmittance for various angles of 0º,10º,20º,30º,40º for c)p =670nm d) p =530nm e) Normalized wavelength shift as a function incident angle. The cyan curve represents a metasurface without a fin with a notch period p= 620nm with a Q of 1165. The light violet curve represents the metasurface with fins with notch period p = 530nm with Q=872. The light blue curve represents the metasurface with fins with notch period p = 670nm. Inset: Schematic of metasurface with fins (blue is silicon and light grey here is sapphire).

### Discussion

We have demonstrated that the resolution and Q factor of a phase gradient metasurface can be disassociated, allowing diffraction to be perfectly sculpted over a large angular range using vanishingly small refractive index biasing. Leading to this advantage, a zero-nearfield-coupling regime was found to result from interference between nearest neighbor interactions mediated by the transverse electric field and those mediated by the longitudinal electric field. Although typically very weak, we enhanced the longitudinal electric field by filling the space between meta-atoms with anisotropic nanofins to balance the two polarizations. As a proof-of-principle for near perfect diffraction shaping, we have numerically demonstrated beam splitting and beam steering covering a wide field of view >106º, enabled by subwavelength spatial resolution of $\lambda/1.6$. Importantly, despite this very fine pixel pitch, diffraction efficiency of over 90% is maintained with a refractive index contrast across the metasurface as low as $2\times10^{-6}$, made possible by the meta-atoms supporting Q as high as 2.8 million. To contextualize the extreme refractive index sensitivity represented by these simulations, we consider physical phenomena capable of producing such index changes. Increasing the temperature of silicon by approximately 11 mK or doping silicon with $1.33 \times 10^{15}$ carriers/cm$^3$, can increase its index by $2 \times 10^{-6}$. In terms of electrical programming, these values imply that signals of just a few mV would be sufficient for switching. Using a coupled dipole model, we also show that near perfect diffraction can be sustained for arbitrarily high Q by working arbitrarily close to the configuration/wavelength where the nearest neighbor coupling coefficient is guaranteed to flip sign. Having identified that access to all of the benefits of decoupling hinges on the existence of a hybrid mode crossing, we extended high Q wave shaping to the full 180º field of view by finding a crossing for a meta-structure with resolution of $\lambda/2.2$. Looking for experimental evidence of decoupling in a uniform metasurface, we chose to focus on incident angle dispersion. As neighboring meta-atoms see an incident wave arriving with a phase difference that grows with incident angle, direct nearfield coupling produces interference that translates a tilted phase front into a resonant shift. To capitalize on this correlation, we measured incident angle dependent transmission spectra for a series of nanofabricated, period swept, metasurfaces. Not only did we find a particular sample with Q>870 that keeps its resonant wavelength within the half linewidth over incident angles ranging from -20° to +20°, but we also observe a transition from red shifting to blue shifting. This transition, that can be explained as a sign flip in the coupling coefficient, serves as a signature for perfect decoupling, since the coupling strength must pass through zero in order to change sign. Our platform provides a blueprint for dense phase gradient metasurfaces with efficiently yet independently addressable meta-atoms, paving the way for high speed, low power, wavefront shaping technologies such as LIDAR, and VR/AR. Beyond programmable wavefront shaping, our approach can be generalized to improve nonlinear and nonreciprocal metasurfaces which also rely on especially weak material coefficients, as well as technologies like multiplexed biosensors and laser transmitters, seeking maximally dense resonator arrays. Finally, the lessons we have learnt are not specific to DGMR meta-atoms and would naturally extend perfect decoupling to other types of optical resonators and other types of resonator lattices.

## Methods

Computational Design

COMSOL Multiphysics 6.1 based on finite-element-method (FEM) was used to perform numerical simulations of our realized structures with periodic boundary conditions in the x and y directions. Port boundary conditions were used in the scattering simulations with the incident plane wave coming from the positive z direction. The Eigenmode simulations were performed with scattering boundary conditions in the z directions.

Fabrication

The metasurfaces were fabricated using standard electron beam (e-beam) lithography. Initially, the silicon-on-sapphire (SOS) wafers (University Wafer) underwent a cleaning process involving sonication in acetone and isopropyl alcohol. Following this, a ZEP520a (ZEON Corporation) film was uniformly formed on the clean SOS surface by spin-coating and subsequently baked for 3 minutes at 180°C. After this baking step, the mask design was created using a 50kV electron beam in an Elionix ELS-S50 system and then developed for 60 seconds in ZEDN50 (ZEON Corporation). This developed mask was used in the standard dry etching process using $SF_6$, $C_4F_8$, and Ar chemistries (Oxford Instrument Plasma Lab 100 System), enabling the transfer of the pattern to the 350 nm thick Silicon layer on the SOS wafer. Once the Si metasurfaces were obtained, the residual mask was removed using Remover PG (Kayaku Advanced Materials, Inc.) heated to 60 °C, followed by a comprehensive cleaning with $O_2$ plasma ashing (Plasma Etch, PE 50 Asher).

Optical Characterization

The transmission spectra were measured with a home-built Fourier plane microscope that included a broadband NKT supercontinuum laser, lenses, polarizers, and a NIR detector (NIRvana camera). An aspheric lens (NA=0.42) is used as an objective lens, collimating the transmitted light from the sample at the focal plane. The sample was placed on a manual rotation stage attached to a kinematic XYZ translational stage(Thorlabs Inc.) which could be manually rotated to change the incidence angle (Supplementary Figure 5) and then record the transmitted intensity. The recorded transmitted intensities are normalized with the transmittance intensity of a bare silicon-on-sapphire substrate. Throughout this paper, the Q-factor was obtained by fitting the spectral data with the function.

$$T = \left| \frac{1}{1 + F sin^2(n_s k h_s)} \right| \left| a_r + a_i i + \frac{b}{f - f_0 + \gamma i} \right|^2$$

The first term of the above expression shows the Fabry-Perot resonances through the substrate with an index of refraction $n_s$ and thickness $h_s$. F is the reflectivity of the air-substrate interface and the free space wavevector of the incoming wave represented by k. The second term is a superposition of the background, $a_r + a_i i$ and the Lorentzian resonance with a resonant frequency $f_0$ and fullwidth at half-maximum of $2\gamma$. Q is calculated as $f_0/2\gamma$.

**Reference**

1. Kuznetsov, A. I. *et al.* Roadmap for Optical Metasurfaces. *ACS Photonics* **11**, 816–865 (2024).

2. Dorrah, A. H. & Capasso, F. Tunable structured light with flat optics. *Science* vol. 376 Preprint at https://doi.org/10.1126/science.abi6860 (2022).

3. Koenderink, A. F., Alù, A. & Polman, A. *Nanophotonics: Shrinking Light-Based Technology*. www.sciencemag.org.

4. Savage, N. Digital spatial light modulators. *Nat Photonics* 170–172 (2009).

5. Staude, I. *et al.* Tailoring directional scattering through magnetic and electric resonances in subwavelength silicon nanodisks. *ACS Nano* **7**, 7824–7832 (2013).

6. Juliano Martins, R. *et al.* Metasurface-enhanced light detection and ranging technology. *Nat Commun* **13**, (2022).

7. Kuznetsov, A. I., Miroshnichenko, A. E., Brongersma, M. L., Kivshar, Y. S. & Luk'yanchuk, B. Optically resonant dielectric nanostructures. *Science* vol. 354 Preprint at https://doi.org/10.1126/science.aag2472 (2016).

8. Yao, K. & Liu, Y. Controlling Electric and Magnetic Resonances for Ultracompact Nanoantennas with Tunable Directionality. *ACS Photonics* **3**, 953–963 (2016).

9. Intaravanne, Y. *et al.* Phase Manipulation-Based Polarization Profile Realization and Hybrid Holograms Using Geometric Metasurface. *Adv Photonics Res* **2**, (2021).

10. Lin, D., Fan, P., Hasman, E. & Brongersma, M. L. Dielectric gradient metasurface optical elements. *Science (1979)* **345**, 298–302 (2014).

11. Raeker, B. O. *et al.* All-Dielectric Meta-Optics for High-Efficiency Independent Amplitude and Phase Manipulation. *Adv Photonics Res* **3**, (2022).

12. Lin, S., Chen, Y. & Wong, Z. J. High-performance optical beam steering with nanophotonics. *Nanophotonics* vol. 11 2617–2638 Preprint at https://doi.org/10.1515/nanoph-2021-0805 (2022).

13. Zheng, G. *et al.* Metasurface holograms reaching 80% efficiency. *Nat Nanotechnol* **10**, 308–312 (2015).

14. Xu, B. *et al.* Metalens-integrated compact imaging devices for wide-field microscopy. *Advanced Photonics* **2**, (2020).

15. Lee, G. Y. *et al.* Metasurface eyepiece for augmented reality. *Nature Communications 2018 9:1* **9**, 1–10 (2018).

16. Solntsev, A. S., Agarwal, G. S. & Kivshar, Y. Y. Metasurfaces for quantum photonics. *Nature Photonics* vol. 15 Preprint at https://doi.org/10.1038/s41566-021-00793-z (2021).

17. Iwanaga, M., Hironaka, T., Ikeda, N., Sugasawa, T. & Takekoshi, K. Metasurface Biosensors Enabling Single-Molecule Sensing of Cell-Free DNA. *Nano Lett* **23**, (2023).

18. Yang, Y. *et al.* Integrated metasurfaces for re-envisioning a near-future disruptive optical platform. *Light: Science and Applications* vol. 12 Preprint at https://doi.org/10.1038/s41377-023-01169-4 (2023).

19. Zhao, Q., Zhou, J., Zhang, F. & Lippens, D. Mie resonance-based dielectric metamaterials. *Materials Today* vol. 12 Preprint at https://doi.org/10.1016/S1369-7021(09)70318-9 (2009).

20. Yao, Y. *et al.* Electrically tunable metasurface perfect absorbers for ultrathin mid-infrared optical modulators. *Nano Lett* **14**, (2014).

21. Yang, Y., Kravchenko, I. I., Briggs, D. P. & Valentine, J. All-dielectric metasurface analogue of electromagnetically induced transparency. *Nat Commun* **5**, (2014).

22. Karvounis, A., Gholipour, B., MacDonald, K. F. & Zheludev, N. I. All-dielectric phase-change reconfigurable metasurface. *Appl Phys Lett* **109**, 051103 (2016).

23. Wang, S. S. & Magnusson, R. *Theory and Applications of Guided-Mode Resonance Filters*. (1993).

24. Barton, D. *et al.* High-Q nanophotonics: Sculpting wavefronts with slow light. *Nanophotonics* **10**, (2020).

25. Horie, Y., Arbabi, A., Arbabi, E., Kamali, S. M. & Faraon, A. High-Speed, Phase-Dominant Spatial Light Modulation with Silicon-Based Active Resonant Antennas. *ACS Photonics* **5**, (2018).

26. Benea-Chelmus, I. C. *et al.* Gigahertz free-space electro-optic modulators based on Mie resonances. *Nat Commun* **13**, (2022).

27. Lawrence, M., Barton, D. R. & Dionne, J. A. Nonreciprocal Flat Optics with Silicon Metasurfaces. *Nano Lett* **18**, 1104–1109 (2018).

28. Hail, C. U., Michaeli, L. & Atwater, H. A. Third Harmonic Generation Enhancement and Wavefront Control Using a Local High-Q Metasurface. *Nano Lett* **24**, 2257–2263 (2024).

29. Shafirin, P. A., Zubyuk, V. V., Fedyanin, A. A. & Shcherbakov, M. R. Nonlinear response of Q-boosting metasurfaces beyond the time-bandwidth limit. *Nanophotonics* **11**, (2022).

30. Qin, H., Redjem, W. & Kante, B. Tunable and enhanced optical force with bound state in the continuum. *Opt Lett* **47**, 1774–1777 (2022).

31. Zhuang, R., He, J., Qi, Y. & Li, Y. High-Q Thin-Film Lithium Niobate Microrings Fabricated with Wet Etching. *Advanced Materials* **35**, 2208113 (2023).

32. Song, W. *et al.* High-density waveguide superlattices with low crosstalk. *Nat Commun* **6**, (2015).

33. Jahani, S. & Jacob, Z. Transparent subdiffraction optics: nanoscale light confinement without metal. *Optica, Vol. 1, Issue 2, pp. 96-100* **1**, 96–100 (2014).

34. Zhang, J. *et al.* Plasmonic metasurfaces with 42.3% transmission efficiency in the visible. *Light Sci Appl* **8**, (2019).

35. Tseng, E. *et al.* Neural nano-optics for high-quality thin lens imaging. *Nature Communications 2021 12:1* **12**, 1–7 (2021).

36. Lee, G. Y., Sung, J. & Lee, B. Recent advances in metasurface hologram technologies. in *ETRI Journal* vol. 41 10–22 (John Wiley and Sons Inc., 2019).

37. Lin, L., Hu, J., Dagli, S., Dionne, J. A. & Lawrence, M. Universal Narrowband Wavefront Shaping with High Quality Factor Meta-Reflect-Arrays. *Nano Lett* (2022) doi:10.1021/acs.nanolett.2c04621.

38. Joannopoulos, J. D., Johnson, S. G., Winn, J. N. W. & Meade, R. D. *Photonic Crystals*. (2007).

39. Koshelev, K., Lepeshov, S., Liu, M., Bogdanov, A. & Kivshar, Y. Asymmetric Metasurfaces with High- Q Resonances Governed by Bound States in the Continuum. *Phys Rev Lett* **121**, (2018).

40. Campione, S. *et al.* Broken Symmetry Dielectric Resonators for High Quality Factor Fano Metasurfaces. *ACS Photonics* **3**, 2362–2367 (2016).

41. Klopfer, E., Delgado, H. C., Dagli, S., Lawrence, M. & Dionne, J. A. A thermally controlled high-Q metasurface lens. *Appl Phys Lett* **122**, (2023).

42. Weiss, A. *et al.* Tunable Metasurface Using Thin-Film Lithium Niobate in the Telecom Regime. *ACS Photonics* **9**, (2022).

43. Vabishchevich, P. & Kivshar, Y. Nonlinear photonics with metasurfaces. *Photonics Res* **11**, (2023).

44. Klopfer, E., Dagli, S., Barton, D., Lawrence, M. & Dionne, J. A. High-Quality-Factor Silicon-on-Lithium Niobate Metasurfaces for Electro-optically Reconfigurable Wavefront Shaping. *Nano Lett* **22**, 1703–1709 (2022).

45. Huang, W.-P. *Coupled-Mode Theory for Optical Waveguides: An Overview*. (1994).

46. Jahani, S. *et al.* Controlling evanescent waves using silicon photonic all-dielectric metamaterials for dense integration. *Nat Commun* **9**, (2018).

47. Mia, M. B. *et al.* Exceptional coupling in photonic anisotropic metamaterials for extremely low waveguide crosstalk. *Optica* **7**, 881 (2020).

48. Sci, M. *Mark Lawrence Symmetry and Topology at the Metasurface*. (2015).

49. Goodman, J. W. Introduction to Fourier Optics 2nd edition. *Book* (1996).

50. Yu, N. *et al.* Light propagation with phase discontinuities: Generalized laws of reflection and refraction. *Science (1979)* **334**, 333–337 (2011).

51. Zhang, X. *et al.* Controlling angular dispersions in optical metasurfaces. *Light Sci Appl* **9**, (2020).

## Acknowledgements

We thank Dr Jennifer Dionne for her insightful discussions. For the use of the equipment and staff assistance, we gratefully acknowledge the financial support of Washington University in St. Louis and the Institute of Materials Science and Engineering. We also thank the Optica Foundation for its financial support.

## Author Contributions

S.A. and L.L. contributed equally to this work. M.L, L.L., B.Z. conceived and designed the experiments. S.A. performed theoretical calculations and numerical simulations. L.L., B.Z., and H.C.D. fabricated the device. S.A., L.L., B.Z. conducted optical characterization of fabricated samples. S.A., L.L., and M.L. wrote the manuscript. ML conceived the idea and supervised the project. All authors discussed the results and commented on the manuscript.

## Competing Interests

The authors declare no competing interests.

## Materials & Correspondence

Correspondence and materials should be addressed to S.A. and M.L.

## Data availability

The data that support the plots within this paper and other findings of this study are available from the corresponding authors upon reasonable request.

# Supplementary Information: Eliminating nearfield coupling in dense high quality factor phase gradient metasurfaces

*Samuel Ameyaw[1][†][*], Lin Lin[1,2][†], Bo Zhao[1], Hamish Carr Delgado[3] and Mark Lawrence[1][*]*

[1]Department of Electrical and Systems Engineering, Washington University in St. Louis, St. Louis, Missouri 63130, USA

[2]Department of Chemistry, Washington University in St. Louis, St. Louis, Missouri 63130, USA

[3]Department of Materials Science and Engineering, Stanford University, Stanford, California 94305, USA

*Corresponding author(s). E-mail(s): ameyaw@wustl.edu ; markl@wustl.edu

†These authors contribute equally

**SUPPLEMENTARY NOTE 1: Beam-steering metasurfaces away from the crossing**

In Figure 1c of the main text, we demonstrate efficient beam steering with a diffraction efficiency of 90% at a refractive index change ($\Delta n$) of $7 \times 10^{-6}$. Here we further elucidate that, away from the crossing point, the efficiency of diffracting light into the +1$^{st}$ diffraction order significantly decreases. This reduction is attributed to the deviation from the zero-coupling regime. For the identical structure and $\Delta n$ value, the diffraction efficiency remarkably drops to 13%, as illustrated in Supplementary Figure 1. This decrease is primarily due to the substantial residual coupling observed within the system.

**SUPPLEMENTARY NOTE 2: Beam-splitting metasurfaces away from the crossing**

In Figures 3c and 3d of the main text, we apply alternate refractive index biasing ($-\Delta n/2$, $+\Delta n/2$) to the Si nanoantennas operating within the zero-coupling regime. This approach achieves near perfect diffraction efficiency in splitting of the incident wave into -1$^{st}$ and +1$^{st}$ diffraction orders with and high quality (Q) factors. Notably, despite the implementation of anisotropic fins to facilitate the fast decay of the evanescent field, residual coupling remains in the designed system when operating away from the zero-coupling regime. This residual coupling constrains the attainable Q factor, which is critical for achieving high diffraction efficiency. In Supplementary Figure 2a, a metasurface operating beyond the crossing is shown to exhibit a Q factor of 197,946 alongside a remarkable diffraction efficiency of 98%. However, when the Q factor is increased to 497,560, the diffraction efficiency is significantly reduced to 53%, underscoring the significant impact of coupling on performance of such systems.

## SUPPLEMENTARY NOTE 3: Details about analytical methods.

To verify the agreement between numerical simulations and the analytical model, we employ the Lorentz resonance model for two coupled dipoles, which is expressed as

$$\begin{bmatrix} \delta_1 + i\gamma_1 & g \\ g & \delta_2 + i\gamma_2 \end{bmatrix} \begin{bmatrix} p_1 \\ p_2 \end{bmatrix} = \begin{bmatrix} g_1 E_{inc1} \\ g_2 E_{inc2} \end{bmatrix} \quad (1)$$

For the matrix above, it is assumed that the incident wave approaches resonance, with $\delta_{1,2} = \omega - \alpha_{1,2} \approx 0$. Taking into account the real parts of the diagonal terms, the multiplicative 2x2 matrix can be written as:

$$\begin{bmatrix} \delta_1 & g \\ g & \delta_2 \end{bmatrix} \quad (2)$$

where $\delta_{1,2}$ are the frequency terms of the coupled dipoles. The 2x2 matrix can further be expanded to include various components, accounting for the frequency contribution from the average refractive indexes $(n_{si} + \delta_{no})$ and the refractive index bias $(\Delta n)$.

$$\begin{bmatrix} \omega_o(n_{si} + \delta_{no}) + \omega_1\left(\frac{\Delta n}{2}\right) & g \\ g & \omega_o(n_{si} + \delta_{no}) - \omega_1\left(\frac{\Delta n}{2}\right) \end{bmatrix} \quad (3)$$

We then calculate the eigenvalues of the above matrix, which correspond to the eigenfrequencies of the simulated coupled nanoantennas with a refractive index bias $(\Delta n)$.

$$\begin{vmatrix} \omega_o(n_{si} + \delta_{no}) + \omega_1\left(\frac{\Delta n}{2}\right) - \lambda & g \\ g & \omega_o(n_{si} + \delta_{no}) - \omega_1\left(\frac{\Delta n}{2}\right) - \lambda \end{vmatrix} = 0 \quad (4)$$

$$\lambda^2 - 2\lambda\big(\omega_o(n_{si} + \delta_{no})\big) + \big(\omega_o(n_{si} + \delta_{no})\big)^2 - \omega_1\left(\frac{\Delta n}{2}\right)^2 - g^2 = 0 \quad (5)$$

Solving the above equation results in the following eigenvalues.

$$\lambda = \omega_o(n_{si} + \delta_{no}) \pm \sqrt{\omega_1\left(\frac{\Delta n}{2}\right)^2 + g^2} \quad (6)$$

$\lambda$ from the above equation is the eigenfrequencies of the two coupled dipoles with refractive index bias $\Delta n$.

When $\Delta n = 0$, for clarity let $\lambda = \lambda'$, and equation (6) simplifies to:

$$\lambda' = \omega_o(n_{si} + \delta_{no}) \pm g \quad (7)$$

From equation (7), the following relations can be derived:

$$\omega_o(n_{si} + \delta_{no}) = \frac{\lambda'_1 + \lambda'_2}{2} \qquad (8)$$

$$g = \frac{\lambda'_1 - \lambda'_2}{2} \qquad (9)$$

Also, the diffraction properties of the dipoles were modeled using complex Fourier series:

$$c_n = \frac{1}{2T}\int_{-T}^{T} f(t) e^{-\frac{i\pi n t}{T}} dt \qquad (10)$$

Considering n = 0, 1, −1, which represent the $0^{th}$, $+1^{st}$, and $-1^{st}$ diffraction orders respectively.

$$c_0 = \frac{1}{2T}\int_{-T}^{0} p_1 dt + \frac{1}{2T}\int_{0}^{T} p_2 dt \qquad (11)$$

$$c_1 = \frac{1}{2T}\int_{-T}^{0} p_1 e^{-\frac{i\pi n t}{T}} dt + \frac{1}{2T}\int_{0}^{T} p_2 e^{-\frac{i\pi n t}{T}} dt \qquad (12)$$

$$c_{-1} = \frac{1}{2T}\int_{-T}^{0} p_1 e^{\frac{i\pi n t}{T}} dt + \frac{1}{2T}\int_{0}^{T} p_2 e^{\frac{i\pi n t}{T}} dt \qquad (13)$$

$$Diffraction\ Efficiency = \frac{|c_1|^2 + |c_{-1}|^2}{|c_0|^2 + |c_1|^2 + |c_{-1}|^2} \qquad (15)$$

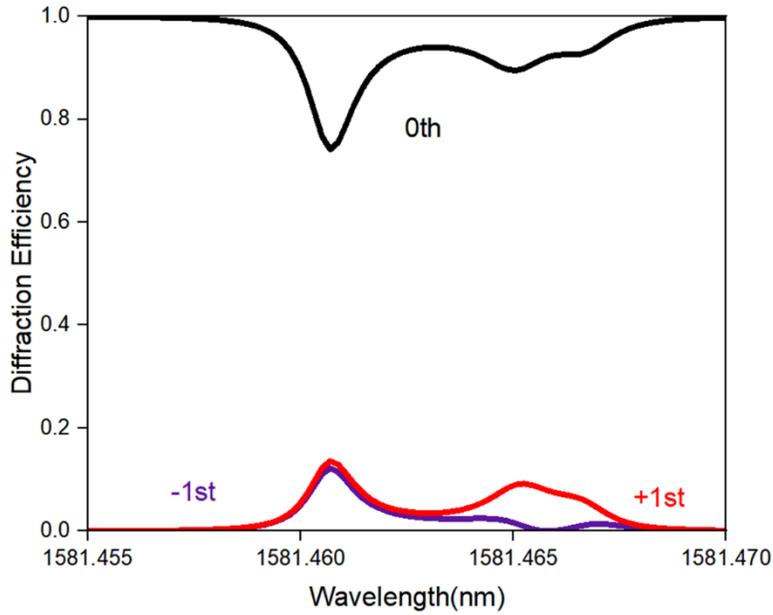

**Supplementary Figure 1.** Diffraction spectrum of metasurfaces with anisotropic fins operating away from the crossing point, $\Delta n=7\times 10^{-6}$, and notch dimensions of 10nm x 30nm, demonstrating a 13% diffraction efficiency.

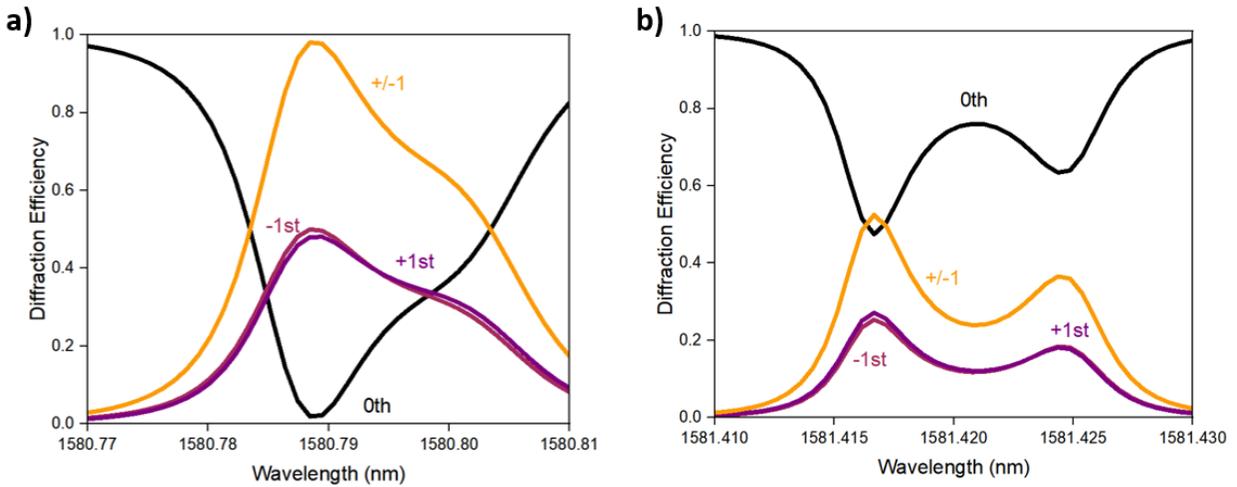

**Supplementary Figure 2.** Diffraction spectra of metasurface beam-splitters away from the crossing, demonstrating (a) diffraction efficiency of 98% with a Q of 197,946, $\Delta n=4\times 10^{-5}$ with notch dimensions of 4nmx100nm, and (b) diffraction efficiency of 53% with a Q of 497,560, $\Delta n=1.45\times 10^{-5}$ with notch dimensions of 10nmx30nm.

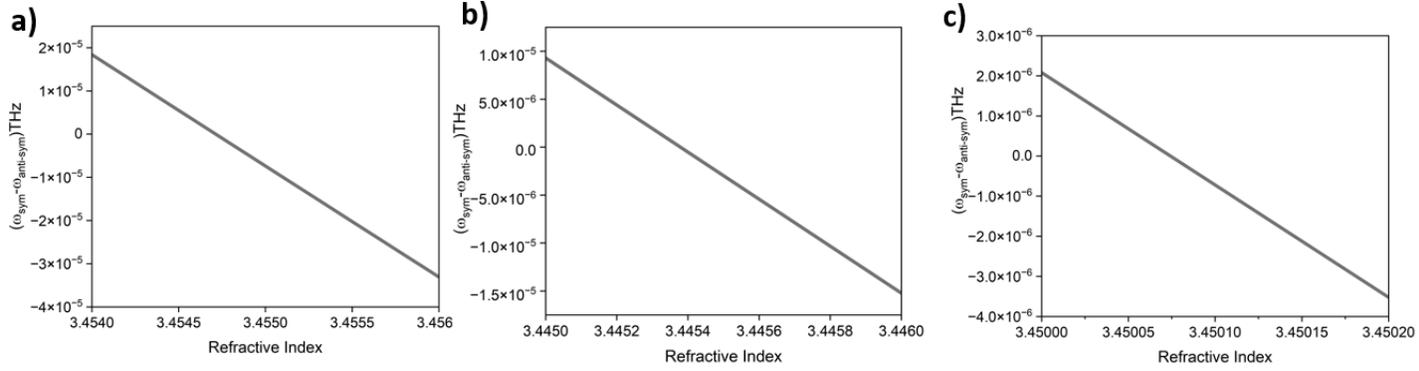

**Supplementary Figure 3.** Plots showing that a crossing is found for the highly efficient wave-shaping devices in the main text. (a) The refractive index used for the Si nanoantennas $n_{si}$=3.454713695 for Figure 1b and 1c in the main text. (b) The refractive index used for the Si nanoantennas $n_{si}$= 3.44537876741189 for Figure 2a in the main text (c) The refractive index used for the Si nanoantennas $n_{si}$= 3.450074282 for Figure 2b in the main text.

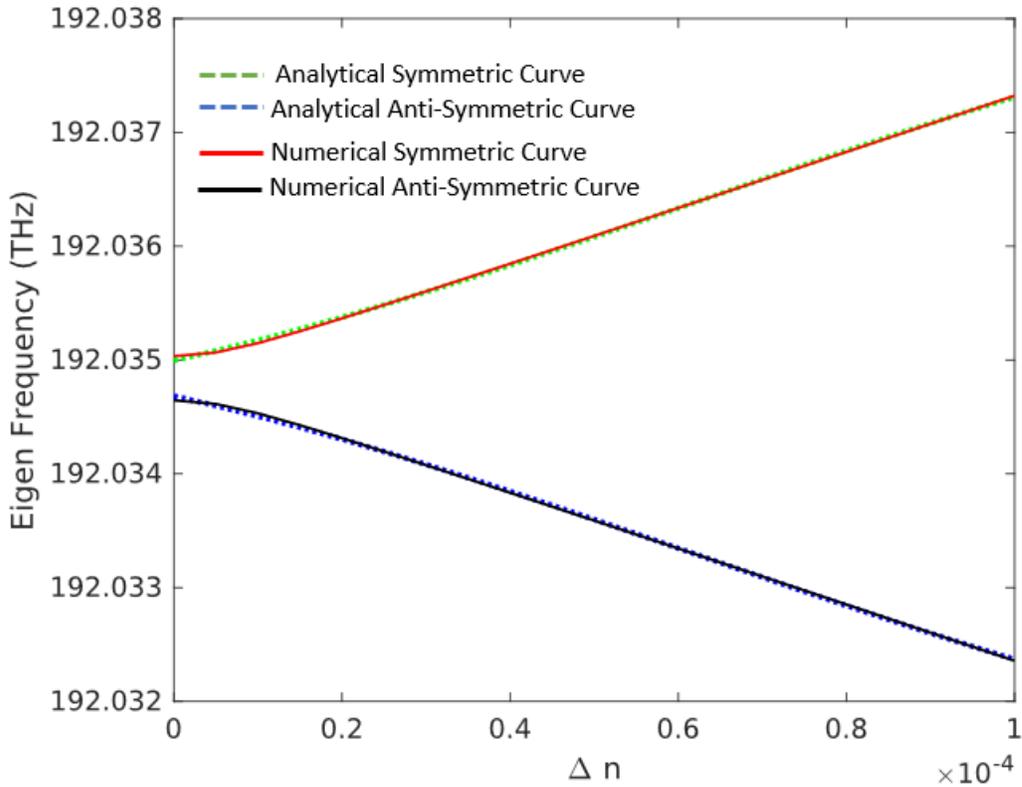

**Supplementary Figure 4:** Comparison of the splitting curves obtained from analytical model and numerical simulation for coupled Si Nanoantennas. In the simulation, an eigenmode simulation of two coupled fin-isolated Si nanoantennas with a refractive index of 3.43 with notch dimension of 10 nm x 40 nm is performed. Then a parametric sweep of the refractive index bias $\Delta n$ is obtained. In the analytical modeling, to get $\omega_o(n_{si} + \delta_{no})$ and g, we set $\Delta n$=0, retrieve the eigenfrequencies, and then inset in equations (8) and (9) to get the parameters respectively.

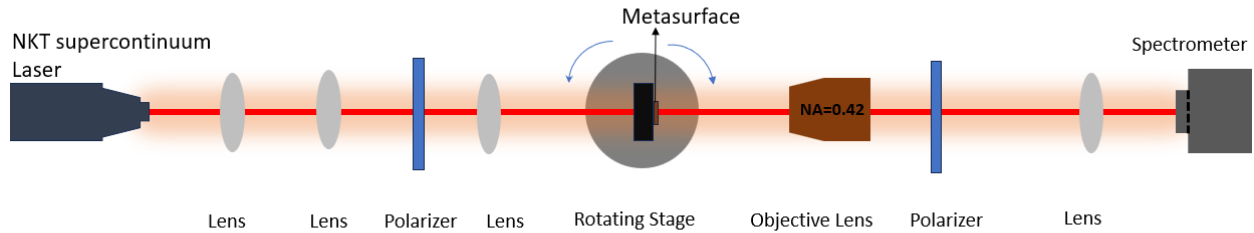

**Supplementary Figure 5:** Schematic of a home-built microscope setup for measuring angular-dependent transmission spectrum.

Reference

1. Mia, M. B. *et al.* Exceptional coupling in photonic anisotropic metamaterials for extremely low waveguide crosstalk. *Optica* **7**, 881 (2020).
2. Sci, M. *Mark Lawrence Symmetry and Topology at the Metasurface*. (2015).
3. Goodman, J. W. Introduction to Fourier Optics 2nd edition. *Book* (1996).